# Development of a Robust Depth-Pressure Estimation Algorithm for a Vision-Based Breast Self-Examination Guidance System

J.A.Jose[1], P.Ching[2], and M. Cabatuan[1]
[1]Department of Electronics and Communications Engineering, De La Salle University, Manila, Philippines.
[2]Department of Industrial Engineering, De La Salle University, Manila, Philippines.
john.anthony.jose@dlsu.edu.ph

*Abstract*—In the case of breast cancer, as with most cancers, early detection can significantly improve a person's chances of survival. This makes it important for there to be an effective and accessible means of regularly checking for manifestations of the disease. A vision-based guidance system (VBGS) for breast self-examination (BSE) is one way to improve a person's ability to detect the cancerous systems. In response to this need, this study sought to develop a depth-pressure estimation algorithm for the proposed VBGS. A large number of BSE videos were used to train the model, and these samples were segmented according to breast size, which was found to be a differentiation factor in the depth-pressure estimation. The result was an algorithm that was applicable for universal use. In addition to these, several feature extraction schemes were tested with the objective of making the algorithm functional on average technology. It was found that Law's Textures Histogram and Local Binary Pattern Global Histogram were the most effective in estimating pressure using visual data. Moreover, combinations of the two schemes further improved the accuracy of the model in estimation. The resulting algorithm was thereby fit to be used by the average consumer.

## I. INTRODUCTION

With breast cancer, as with most cancers, early detection can significantly improve a person's chances of survival. Studies have shown that the major difference between survival rates in developed countries may be attributed to early detection [1]. This makes it important for people to have access to convenient means of detecting and identifying symptoms of the disease.

For developing countries, such as the Philippines, where few people have access to medical care, the tests that are designed for detecting breast cancer may be out of reach for most. This leaves Breast Self-Examination (BSE), wherein instead of paid medical professional conducting the breast examination, the patient conducts the breast examination on herself. Varying success rates have been reported for BSE [2]–[4]. These differences may depend on the extent of enforcement and instruction being given to the populations that had supposedly been applying BSE. By allowing the participants to conduct the exam on their own, much is left to chance on their level of understanding of the instructions and the frequency at which they conduct BSE [5]. For the former, at least, some form of guidance may be provided through the use of vision-based guidance systems which can identify and correct deviation from the correct BSE procedure[5]–[7].

### A. Computer Vision in Breast Cancer Detection

As of late, computer vision has been widely applied as an alternative low-cost solution for developing countries even in different fields [8], [9]. The majority of computer vision applications directed at breast cancer detection make use of the results of medical exams such as mammograms. Although mammograms have reduced breast cancer mortality rates by fostering early detection, breast cancer continues to be the leading cause of cancerous death among women [6]. Human limitations on the part of the medical professionals interpreting the results are hypothesized to limit the effectiveness of mammograms, to some extent. Hence, a number of computer aided diagnosis systems have been developed and are continuously being improved to augment weaknesses in this aspect [10].

Where previous computer aided diagnosis systems tended to have a high rate of false positive ROI detection, a recent study has successfully addressed this gap with the use of feature extraction schemes in describing the mammographic regions, and testing various classifiers [11]. This framework showed significantly better results in accurately detecting lesions, as compared to older frameworks that were limited to identifying ROIs containing supposed masses. Among the classifiers tested, random forest was found to have the highest accuracy rate (81.09%) followed closely by SVM (80.01%).

Another study tackled the selection of parameters being used in the existing lesion detection and segmentation models, thereby improving their efficiency[12]. Hitherto, these parameters were typically selected through manual search techniques, which were known to be time-consuming and sub-optimal. The results exhibited greater capability in identifying multiple ranges of solutions, as well as greater segmentation accuracy on the whole.

However, it is worthy of note that while advancements in computer aided detections systems for medical professionals are undoubtedly conducive towards the early detection of breast cancer, these are only effective after the patient herself becomes aware that she has symptoms of the disease. As the patient may remain ignorant of the symptoms for some time after they manifest, having highly effective guidance systems past this point may be superfluous. On the other hand, there is still great room for improvement in the development of guidance systems to be used by the patients in checking themselves for symptoms.

*B. Vision Based Guidance in BSE*

A number of vision-based guidance systems (VBGS) have been developed in response to the difficulties experienced by people in properly executing BSE. Such systems aim to capture, evaluate, and provide tailored feedback for individual BSE cases.

These systems mainly vary in the form and purpose of the feedback that they provide. A research group called BIOCORE has developed a system that evaluates pre-recorded videos of users conducting BSE. These videos are captured using 6 video cameras recording in parallel, to create an integrated reality system (iRis) that would allow users to review their BSE procedure from all angles [13]. Guidance is thus provided in the form of a virtual reality (VR) experience; errors in the BSE procedure are highlighted and explained by the VBGS [14], and users can better understand the nature of their errors by viewing the video from different angles. The resulting video playback and VR experience is certainly conducive to deeply understanding the correct BSE procedure; however, the effectiveness of this approach as a guidance system for BSE would depend on the actual needs of the user.

The trend in VBGS has been leaning towards providing real-time instruction and correction of errors, as opposed to a detailed evaluation of the information that was captured [4]. This trend is also evident in recent VBGS developments for BSE. The VBGS model designed by Hu et al. [15] can recognize a breast image and present the user with a delineation of the breast area in the image. The image serves as instruction for the user on the areas that should be palpated. As the user follows these instructions, error detection can happen in real time through hand motion recognition. While the method is still not fully robust due to potential variations in the appearance of the breast (i.e. age) and differences in the environment of capture (i.e. lighting), [16] nonetheless shows the potency of computer vision in BSE.

*C. Depth Estimation in Guided BSE*

Successful breast examination is largely dependent on applying the appropriate amount of pressure in the breast to detect a lump or lesion. One means of visually gauging the pressure being applied to the breast is through the change in depth as the hand palpates the breast. Given that there have already been a number of algorithms developed for measuring the depth of objects [17], [18], depth estimation is a promising substitute for measuring pressure. However, at the moment, there have been few attempts to integrate depth estimation in VBGS for BSE.

A research group from Coventry University has developed a means of measuring the depth of breast palpation through entropy [17]. This method compares the difference between two successive images in a video in order to gauge depth, and through it, the amount of pressure applied. However, the resulting depth estimations were not numerically accurate. This would suggest that there are other parameters to consider in depth estimation, aside from the physical displacement of the hand. Another weakness of the method was that it could only consider up-and-down hand movements, when the hand moves in other ways during breast palpation (e.g. translational, rotational).

In a different study, a neural network for depth estimation in breast palpation during BSE was developed. To build its capability in estimating depth, the model was trained using an actual video of BSE with varying levels of depth in breast palpation [18]. This is a step above the previous study, as it can consider palpation in more than one form. However, the robustness of the model in being able to gauge the amount of pressure applied to any person's breasts may still require improvement, as both the testing and training sets were obtained from the same video.

*D. Research Gap*

This study seeks to improve the existing depth estimation algorithms for breast palpitation, with the purpose of contributing to the development of a more accurate and responsive VBGS for BSE [19]. Based on the existing literature on the subject, there is still much to be improved in terms of robustness. The existing algorithms tend to be designed for one type of breast alone and draw data from limited samples. In order to be applicable for a universal VBGS for any user, the depth estimation algorithm must be applicable for all breasts, to the best extent possible. As an additional consideration given that a VBGS for BSE would be used regularly, this study also aims to design the algorithm for ubiquity, in the sense that it should be able to process information using basic technology.

II. METHODS

*A. Dataset Construction*

From the existing studies on depth estimation for BSE, it was established that low robustness of the models being developed came from designing algorithms that were over-fitted to one kind of breast. The current study aims to address this gap, by gathering sufficient visual data on breasts, in order to develop an algorithm that is applicable for a universal BSE guidance system. In particular, sufficient visual data was collected and segmented according to cup size A, B and C, as this may affect depth-pressure conversion. The study recognizes that the exact proportionality of the depth of palpation to pressure may vary across different breast sizes. Displacement from the camera would be smaller for larger breast sizes. Moreover, larger breast sizes may require deeper palpation for the palpation to count as high pressure. The number of samples in the resulting datasets for each quadrant of interest are given in Table 1.

Table 1
Size of Dataset used in Training the Model

| Frame | Cup A | | Cup B | | Cup C | |
|---|---|---|---|---|---|---|
| | Train | Test | Train | Test | Train | Test |
| Left_Q2 | 101 | 18 | 142 | 25 | 61 | 11 |
| Left_Q3 | 120 | 21 | 113 | 19 | 85 | 15 |
| Right_Q2 | 105 | 18 | 120 | 21 | 75 | 13 |
| Right_Q3 | 99 | 18 | 108 | 19 | 81 | 14 |

All visual data was captured in a single conference room in De La Salle University for consistency. An optimal camera-to-user distance of 0.6 to 0.8 m has also been established, both for the sake of consistency and to maximize the torso area captured in the video frame. These videos were captured using a Kinect for Xbox 360, for the reason that the device is capable of extraction video frames as RGB-Depth images. Fig. 1 exhibits the depth estimation capabilities of the Kinect in the depth images (HSV colormaps) shown after each set of sample frames from the actual video. However, the depth of other objects in the frame were also captured, making negligible the differences in depth caused by palpation. It was thus necessary for a specific region of interest to be extracted, such that only depth measures relevant to the pressure of palpation are obtained.

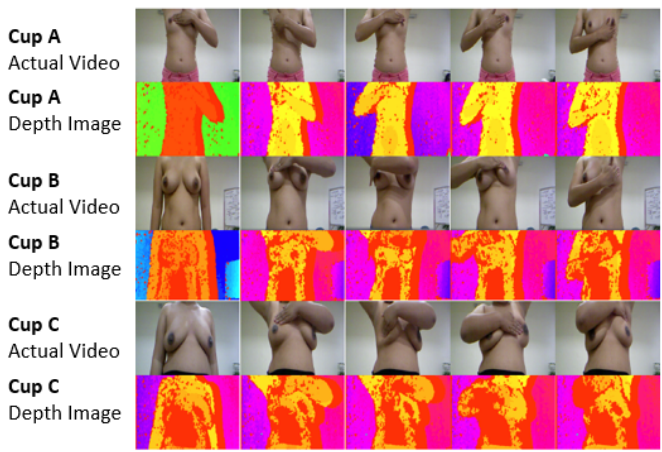

Figure 1: Sample RGB-D sequence recorded via Kinect.

### B. Region of Interest Extraction

The video frames contain two main objects of interest in relation to SBE: the breasts and the fingers used in palpation. To limit the depth information being captured to these objects of interest, mask sequences were created to limit perception. For the breast, perception is limited to the quadrant that is currently being palpated. A rectangular (box) mask sequence was thus created using Adobe Photoshop to capture the quadrant of the breast being palpated. As division of the torso image into quadrants (Fig. 2) would exponentially increase the amount of data to be processed, the dataset of this study was limited to Quadrants 2 and 3 of both the left and right breast.

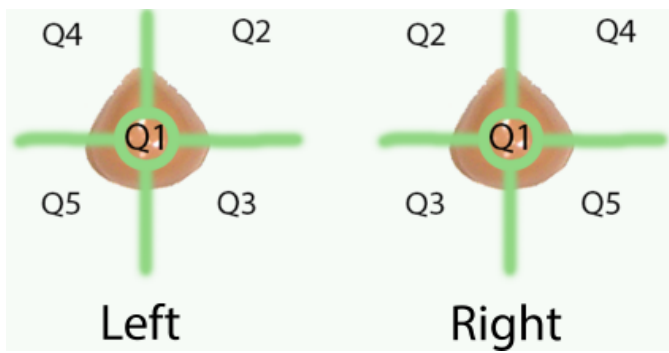

Figure 2: Segmentation of the breast images into quadrants.

On the other hand, extraction of the finger images required the creation of an irregularly-shaped mask sequence. The Roto brush tool of Adobe After Effects CS6 was thus used to create and track a user-identified shape in the video. Said shape was fitted into the appearance of the three fingers used in palpation. When used in conjunction with the box mask sequence, a scalar depth measure for the depth of palpation could then be obtained.

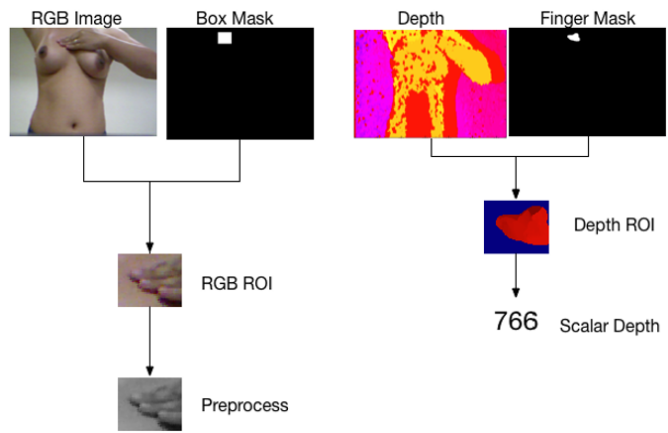

Figure 3: Depth level extraction of the palpation.

### C. Region of Interest Extraction

A fuzzy-based algorithm is used in identifying the corresponding pressure level for each depth measure, although it must be noted that conversion of depth to pressure is not completely continuous. In this algorithm, the range of depth values adjacent to the minimum depth recorded is classified as low pressure, and likewise, the range of depth values adjacent to the maximum depth recorded is classified as high pressure. The "fuzzy-like" nature of the algorithm applies to the range of values between the purely low and high ranges, such that some values are between low and medium, with varying degrees of 'low-ness' and 'medium-ness'; and likewise for the values between medium and high. The A1, A2 and A3 values which determine pressure classification are given in Eqs. 1, 2 and 3.

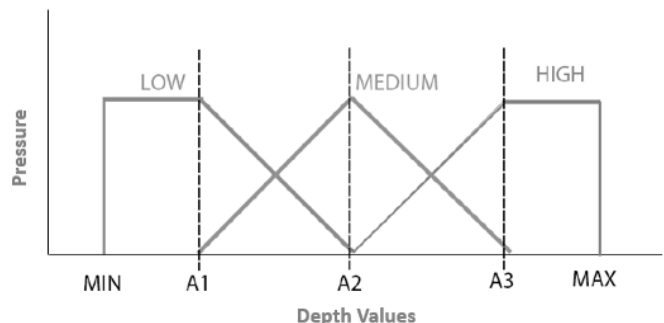

Figure 4: Fuzzy-based depth-pressure conversion.

$$A1 = 0.25 * (MAX - MIN) + MIN \quad (1)$$

$$A2 = 0.50 * (MAX - MIN) + MIN \quad (2)$$

$$A3 = 0.75 * (MAX - MIN) + MIN \quad (3)$$

*Journal of Telecommunication, Electronic and Computer Engineering*

*D. Machine Learning Algorithms*

Four feature extraction schemes were tested for their ability to match the pressure classifications identified using the Kinect depth measurements. Such schemes would allow the VBGS to be usable on even the most basic platforms (i.e. webcam and personal computer), for the sake of accessibility. The four schemes that were tested in this study are: (1) entropy, (2) normalized shadow area, (3) Law's textures histogram, and (4) local binary pattern global histogram.

While each scheme has strengths and weaknesses of its own, its combination with machine learning algorithms would determine its overall capability in depth-pressure estimation. The machine learning algorithms that will be tested in this study are (1) linear regression, (2) Support Vector Machine (SVM), (3) Gradient Boosted Trees (GBT), and (4) Artificial Neural Network (ANN).

## III. RESULTS AND ANALYSIS

*A. Scalar Depth Measure Results*

Table 2 shows the ranges of scalar depth measures that correspond to low, medium and high pressure, according to eq. 1, 2 and 3. These would determine the "correct" conversion of depth to pressure, which the algorithms developed from the feature extraction schemes will be evaluated against.

Table 2
Depth Range for Low, Medium and High Pressure

| Size | Frame | LOW | MEDIUM | HIGH |
|---|---|---|---|---|
| Cup A | Left_Q2 | 762 - 774 | 774 - 782 | 782 – 794 |
|  | Left_Q3 | 744 – 754.1 | 754.1 – 760.9 | 760.9 – 771 |
|  | Right_Q2 | 772 – 778.8 | 778.8 – 783.2 | 783.3 – 790 |
|  | Right_Q3 | 771 – 782.3 | 782.3 – 789.8 | 789.8 – 801 |
| Cup B | Left_Q2 | 607 – 633.6 | 633.6 - 651.4 | 651.4 – 678 |
|  | Left_Q3 | 603 – 608.3 | 608.3 – 611.8 | 611.8 – 617 |
|  | Right_Q2 | 619 – 643.4 | 643.4 – 659.6 | 659.6 – 684 |
|  | Right_Q3 | 614 – 630.5 | 630.5 – 641.5 | 641.5 - 658 |
| Cup C | Left_Q2 | 568 – 583.4 | 583.4 – 593.6 | 563.6 – 609 |
|  | Left_Q3 | 563 – 579.5 | 579.5 – 590.5 | 590.5 – 607 |
|  | Right_Q2 | 591 – 619.9 | 619.9 – 639.1 | 639.1 – 668 |
|  | Right_Q3 | 597 – 625.5 | 625.5 – 644.5 | 644.5 - 673 |

*B. Evaluation of the Machine Learning Algorithm*

For purposes of evaluation, the model developed by Coventry University [14] will be used as the baseline for the algorithm that will be developed in this study. Said model had a success rate of 33.33% in identifying the amount of pressure exerted in palpation. This rate amounts to little more than a random chance of success, based on the segmentation and distribution of data. From the machine learning algorithms that were tested, 3 out of 4 resulted in greater success rates than the baseline; these are SVM, REG, and GBT (see Fig. 5).

It is apparent that GBT was the greatly successful in identifying pressure levels from data included in its training set. However, its performance with the test set, in relation to that with its training set, is symptomatic of overfitting. On the other hand, while methods such as SVM and REG were not as successful as GBT in training, their performance across both training and testing sets is more consistent, indicating that these algorithms may be more capable of accommodating new data outside of the current dataset. With SVM having a highest success rate with its test set, this method will be used in the development of the depth-pressure algorithm.

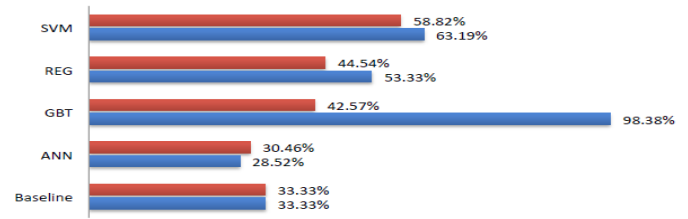

Figure 5: Benchmarking of different machine learning models. Red bar shows testing accuracy while the blue bar shows the training accuracy

*C. Evaluation of Feature Extraction Schemes*

Fig. 6 shows the results of training the SVM model on each of the previously mentioned feature extraction schemes. It is apparent from its performance with both the training and testing sets that Law's textures histogram (Law) and local binary pattern global histogram (LBP) are significantly more effective at identifying the correct pressure levels. With a testing success rate of 67.26% and 70.64%, the performance of the feature extraction schemes is well beyond the performance of the baseline.

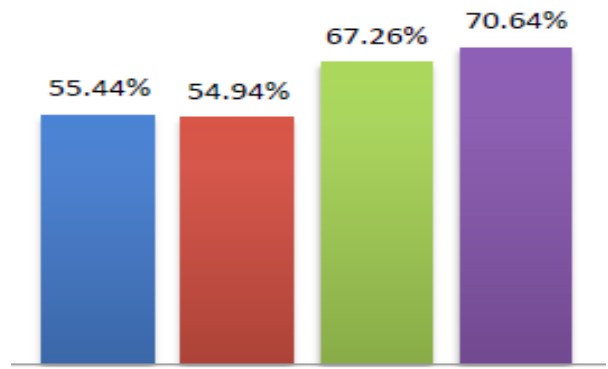

Figure 6: Accuracy of feature extraction schemes. Blue bar represents the entropy features, red bar represents the shadow features, green bar represents laws' textures histogram, and purple bar represents the local binary pattern histogram.

*D. Accuracy Assessment of Combined Feature Extraction Schemes*

To further improve the performance of the model, an attempt is made to combine feature extraction schemes. All combinations which involved the Law and LBP, the two schemes that were most successful independently, resulted in even better performance than independent performance of Law and LBP--with the minor exception of ShaLaw, which is only superior to Law. Thus, it may be concluded that a depth-pressure estimation algorithm composed of the SVM learning algorithm, and the Law and LBP feature extraction schemes, would effectively serve a universal SBE guidance system, capable of being run using the average laptop.

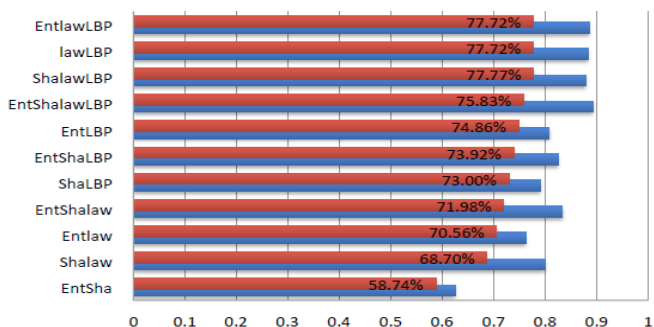

Figure 7: Accuracy of combined feature extraction schemes. Red shows the testing accuracy while blue shows the training accuracy

IV. SUMMARY

The study at hand sought to develop a depth-pressure estimation algorithm for a Breast Self-Examination (BSE) guidance system. In doing so, it aimed to address to major research gaps in this area of study: (1) the absence of a model that could evaluate different kinds of breast, for purposes of mass consumption, and (2) the absence of a model that was capable of being run using average technology, for purposes of greater accessibility.

In order to develop a more robust depth-pressure estimation algorithm for BSE, the study took into account differences in breast size, and differentiated its data accordingly. It was observed that other studies had a tendency to either overfit to a single sample or ignore the differences in the appearances of breasts altogether. This is a severe constraint, given that there are many ways that breasts can vary in appearance. In this respect, the study addressed a significant portion of the gap by segmenting samples according to the major breast cup sizes.

Where ubiquity is concerned, the study made use of feature extraction schemes which would enable the algorithm to be run using an average webcam and personal computer. Out of the four feature extraction schemes, it was found that Law's textures histogram and the local binary pattern global histogram were the most successful at identifying the appropriate pressure levels given a depth image.

Future studies may seek to improve the accuracy of the classifier, as the schemes in this study only made use of simple summary statistics. Options such as an LBP histogram with multiple windows [20]and Dalal-Triggs HOG [21] are largely applicable to the functions of this study. Where depth perception is concerned, it is also surmised that non-texture features (e.g. shade from light, structure from motion) may prove more accurate.

ACKNOWLEDGEMENT

This research was supported by the Department of Science and Technology of the Republic of the Philippines. We would also like to thank our colleagues from the Intelligent Systems Laboratory of De La Salle University – Manila for the support and insight that they contributed to this research.

REFERENCES

[1] BreastCancer.org, "U.S. Breast Cancer." [Online]. Available: http://www.breastcancer.org/symptoms/understand_bc/statistics.

[2] R. K. C. Billones, R. A. A. Masilang, J. A. C. Jose, and E. P. Dadios, "Intelligent operating architecture for audio-visual Breast Self-Examination Multimedia Training System," in *TENCON 2015 - 2015 IEEE Region 10 Conference*, 2015, pp. 1–6.

[3] J. A. C. Jose, M. K. Cabatuan, E. P. Dadios, and L. A. G. Lim, "Stroke position classification in breast self-examination using parallel neural network and wavelet transform," in *TENCON 2014 - 2014 IEEE Region 10 Conference*, 2014, pp. 1–5.

[4] E. Mohammadi *et al.*, "Real-Time Evaluation of Breast Self-Examination Using Computer Vision," *Int. J. Biomed. Imaging*, vol. 2014, 2014.

[5] J. A. C. Jose, M. K. Cabatuan, E. P. Dadios, and L. A. G. Lim, "Depth estimation in monocular Breast Self-Examination image sequence using optical flow," in *2014 International Conference on Humanoid, Nanotechnology, Information Technology, Communication and Control, Environment and Management (HNICEM)*, 2014, pp. 1–6.

[6] R. A. A. Masilang, "Design and Development of a Computer Vision-based Breast Self-Examination Instruction and Supervision System," 2014.

[7] R. A. A. Masilang, M. K. Cabatuan, and E. P. Dadios, "Hand initialization and tracking using a modified KLT tracker for a computer vision-based breast self-examination system," in *Humanoid, Nanotechnology, Information Technology, Communication and Control, Environment and Management (HNICEM), 2014 International Conference on*, 2014, pp. 1–5.

[8] R. A. Bedruz, E. Sybingco, A. Bandala, A. R. Quiros, A. C. Uy, and E. Dadios, "Real-time vehicle detection and tracking using a mean-shift based blob analysis and tracking approach," in *2017IEEE 9th International Conference on Humanoid, Nanotechnology, Information Technology, Communication and Control, Environment and Management (HNICEM)*, 2017, pp. 1–5.

[9] A. R. F. Quiros *et al.*, "A kNN-based approach for the machine vision of character recognition of license plate numbers," in *TENCON 2017 - 2017 IEEE Region 10 Conference*, 2017, pp. 1081–1086.

[10] R. K. C. Billones *et al.*, "Speech-controlled human-computer interface for audio-visual breast self-examination guidance system," in *2015 International Conference on Humanoid, Nanotechnology, Information Technology,Communication and Control, Environment and Management (HNICEM)*, 2015, pp. 1–6.

[11] S. Dhahbi, W. Barhoumi, J. Kurek, B. Swiderski, M. Kruk, and E. Zagrouba, "False-positive reduction in computer-aided mass detection using mammographic texture analysis and classification," *Comput. Methods Programs Biomed.*, vol. 160, pp. 75–83, 2018.

[12] L. Morra, N. Coccia, and T. Cerquitelli, "Optimization of computer aided detection systems: An evolutionary approach," *Expert Syst. Appl.*, vol. 100, pp. 145–156, 2018.

[13] A. Oikonomou and S. Amin, "IRiS: an interactive reality system for breast self-examination training," in *IEEE Engineering in Medicine and Biology Society*, 2004, pp. 5162–5165.

[14] A. Oikonomou, S. Amin, R. N. G. Naguib, A. Todman, and H. AI-Omishy, "Breast Self Examination Training Through the Use of Multimedia:A Prototype Multimedia Application," in *IEEE Engineering in Medicine and Biology*

*Society*, 2003, pp. 1295–1298.
[15] Y. Hu and R. N. G. Naguib, "Search strategies for the automatic delineation of the breast area in a multimedia breast self-examination system," in *IEEE Engineering in Medicine and Biology Society*, 2003, pp. 1315–1318.
[16] J. A. C. Jose, M. K. Cabatuan, R. K. Billones, E. P. Dadios, and L. A. G. Lim, "Monocular depth level estimation for breast self-examination (BSE) using RGBD BSE dataset," in *IEEE Region 10 Annual International Conference, Proceedings/TENCON*, 2016, vol. 2016–Janua.
[17] S. Chen, R. G. Naguib, and A. Oikonomu, "Hand pressure estimation by image entropy for a real-time breast self-examination multimedia system.," in *IEEE Engineering in Medicine and Biology Society*, 2005, pp. 1732–1735.
[18] M. Cabatuan and E. Dadios, "Computer vision-based breast self-examination stroke position and palpation pressure level classification using artificial neural networks and wavelet transforms," in *IEEE Engineering in Medicine and Biology Society*, 2012, pp. 6259–62.
[19] R. K. C. Billones, E. P. Dadios, and E. Sybingco, "Design and Development of an Artificial Intelligent System for Audio-Visual Cancer Breast Self-Examination," *J. Adv. Comput. Intell. Intell. Informatics*, vol. 20, no. 1, pp. 124–131, 2016.
[20] T. Ahonen, a. Hadid, and M. Pietikainen, "Face Description with Local Binary Patterns: Application to Face Recognition," *IEEE Trans. Pattern Anal. Mach. Intell.*, vol. 28, no. 12, pp. 2037–2041, 2006.
[21] N. Dalal and B. Triggs, "Histograms of Oriented Gradients for Human Detection," *2005 IEEE Comput. Soc. Conf. Comput. Vis. Pattern Recognit.*, vol. 1, pp. 886–893, 2005.